\documentstyle{elsart}
\begin{document}
\input{axodraw.sty}
\input{psfig}
\begin{frontmatter}
\title{Radiative Muon Capture by a Proton in Chiral
Perturbation Theory}
\author{T. Meissner\thanksref{mei},}
\address{Department of Physics, Carnegie Mellon University,\\
Pittsburgh, PA 15213, U.S.A.}
\author{F. Myhrer\thanksref{myh}} \and \author{K. Kubodera\thanksref{kub}}
\address{Department of Physics and Astronomy,\\
University of South Carolina,
Columbia, SC 29208, U.S.A.}
\thanks[mei]{email: meissner@yukawa.phys.cmu.edu}
\thanks[myh]{email: myhrer@nuc003.psc.sc.edu}
\thanks[kub]{email: kubodera@nuc003.psc.sc.edu}
\begin{abstract}
The first measurement of the radiative muon capture (RMC) rate
on a proton was recently carried out at TRIUMF.
The TRIUMF group analyzed the RMC rate,
$\Gamma_{\rm RMC}^{exp}$,
in terms of the theoretical formula of Beder and Fearing,
and found the surprising result that 
$g_P\equiv f_P(q^2=-0.88m_\mu^2)$
is 1.5 times the value expected from PCAC. 
To assess the reliability 
of the theoretical framework used by
the TRIUMF group to relate $\Gamma_{\rm RMC}$
to the pseudoscalar form factor $f_P$,
we calculate $\Gamma_{\rm RMC}$ 
in chiral perturbation theory,
which provides a systematic framework 
to describe all the vertices involved in RMC,
fulfilling gauge-invariance 
and chiral-symmetry requirements
in a transparent manner. 
As a first step we present a 
chiral perturbation calculation at tree level
which includes
sub-leading order terms. 
\end{abstract}

\end{frontmatter}

\section{Introduction}

It has long been a great experimental challenge
to observe radiative muon capture (RMC) on the proton, 
$\mu^-+p\rightarrow n+\nu_\mu+\gamma$,
because of its extremely small branching ratio.
Recently, an experimental group 
at TRIUMF \cite{jonetal96} finally succeeded
in measuring $\Gamma_{\mathrm{RMC}}$,
the capture rate for RMC on a proton.\footnote{
To be more precise, the TRIUMF experiment
determined the partial capture rate 
$R(>60\mathrm{MeV})$, corresponding to
emission of a photon with $E_\gamma>60 \mathrm{MeV}$.}
The matrix element of the hadronic charged weak current 
$h^\lambda=V^\lambda -  A^\lambda$
between a proton and a neutron  
is given by 
\begin{eqnarray}
&\langle &
n(p_f) | V^\lambda - A^\lambda | p (p_i) 
\rangle  
\; =  
\nonumber \\
&{\bar {u}}& (p_f) \left[f_V(q^2)\gamma^\lambda
 + \frac{f_M(q^2)}{2m_N}
\sigma^{\lambda\mu}q_\mu
 + f_A(q^2)\gamma^\lambda\gamma_5
 + \frac{f_P(q^2)}{m_\pi}q^\lambda\gamma_5
\right] u(p_i),
\label{eq:formfactors}
\end{eqnarray}
where $q\equiv p_i-p_f$, and 
the absence of the second-class current is assumed.
Of the four form factors appearing 
in eq.(\ref{eq:formfactors}),
$f_P$ is experimentally the least well known.
Although ordinary muon capture (OMC) on a proton, 
$\mu^-+p\rightarrow n+\nu_\mu$,
can in principle give information on $f_P$,
its sensitivity to $f_P$ is intrinsically suppressed.
This is because the momentum transfer involved in OMC,
$q^2=-0.88 m_\mu^2$, 
is far away from the pion-pole position $q^2=m_\pi^2$,
where the contribution of $f_P(q^2)$ 
becomes most important.
RMC on a proton
provides a more sensitive probe of $f_P $ than OMC, 
because the three-body final state in RMC
allows one to come closer to the pion pole. 

To relate $\Gamma_{\mathrm{RMC}}$ to $f_P$,
the authors of \cite{jonetal96} used 
the theoretical framework of Beder and Fearing \cite{bf87}.
In this framework, as in many earlier works
\cite{mw59,opa64,fea80,rt65},
one invokes a minimal substitution 
to generate the RMC transition amplitude 
from the transition amplitude for OMC,
the hadronic part of which is given by Eq. (\ref{eq:formfactors}).
The actual procedure used in \cite{bf87} is as follows.
First, the pion-pole factor is explicitly extracted 
from $f_P$ as $f_P(q^2)=\tilde {f}_P/(q^2-m_\pi^2)$, 
where $\tilde {f}_P$ is a constant.
Then one replaces every $q$ in Eq.(\ref{eq:formfactors})
with $q-e{\cal A}$ 
(${\cal A}$ is the electromagnetic field)
except the $q$ appearing in the $q^2$ dependence 
of $f_V$, $f_A$ and $f_M$.
$\Gamma_{\mathrm{RMC}}^{\mathrm{theor}}$
resulting from this treatment has a parametric dependence
on ${\tilde {f_P}}$. 
In the analysis in ref. \cite{jonetal96},
${\tilde {f_P}}$ is adjusted to optimize
agreement between 
$\Gamma_{\mathrm{RMC}}^{\mathrm{theor}}$ 
and the measured rate 
$\Gamma_{\mathrm{RMC}}^{\mathrm{exp}}$ 
(more precisely $R( > 60 \;{\mathrm{MeV}} )$). 
The result of this optimization,
expressed in terms of 
$g_P\equiv f_P(q^2=-0.88m_\mu^2)
=\tilde{f}_P/(-0.88m_\mu^2-m_\pi^2)$,
is $g_P=(10.0\pm0.9\pm0.3)g_A$,
where $g_A = f_A (0)$.
This value is $\sim 1.5$ times the value
expected from PCAC.
This surprising result should be contrasted 
with the fact that $g_P$ measured in OMC
is consistent with the PCAC prediction
within large experimental uncertainties
\cite{omc}.

A natural question one could ask is:
How reliable is 
$\Gamma_{\mathrm{RMC}}^{\mathrm{theor}}$
used in deducing $g_P$ from 
$\Gamma_{\mathrm{RMC}}^{\mathrm{exp}}$ ?
It seems important to reexamine the reliability
of the existing phenomenological approach
\cite{bf87} which uses a selective minimal substitution.
Chiral perturbation theory (ChPT) provides 
a systematic framework to describe
the electromagnetic-, weak-, and strong-interaction
vertices in a consistent manner,
thereby allowing us to avoid applying 
a phenomenological minimal-coupling substitution
at the level of the transition amplitude. 
Furthermore, ChPT enables us to satisfy 
the gauge-invariance and chiral-symmetry requirements
in a transparent way.

Starting with the seminal work of 
Gasser and Leutwyler \cite{gl84}
ChPT has proven to be a very powerful 
and successful technique for hadronic phenomenology
at low energies \cite{ecker}-\cite{rho}.
Muon capture is another favorable case for applying ChPT
since momentum transfers involved here do not exceed $m_\mu$, 
and $m_\mu$ is small compared to the chiral scale 
$\Lambda \sim $ 1 GeV, indicating the possibility of a reasonably 
rapid convergence of the chiral expansion.
In the case of OMC, Bernard et al.\cite{bkm94} 
and Fearing et al.\cite{feaetal97}
used heavy-baryon ChPT to evaluate $f_P$ 
with better accuracy than achieved 
in the PCAC approach. 
In the case of RMC, a ChPT calculation 
provides a natural extension 
of the classic work of Adler and Dothan\cite{ad66}
based on the low-energy theorems. 
These observations motivate us to attempt
a systematic ChPT calculation of 
$\Gamma_{\mathrm{RMC}}$.
As a first step we calculate 
the total capture rate $\Gamma_{\mathrm{RMC}}$
and the spectrum of the emitted photons, 
$d\Gamma_{\mathrm{RMC}} (k) / dk  $,
to sub-leading order in chiral perturbation expansion.
Thus, our calculation includes nucleon recoil 
contributions of ${\cal{O}} (1/M)$. 
We limit ourselves here to the case
of RMC from the $\mu$-$p$ atom 
with statistical spin distributions,
leaving out the hyperfine-state decomposition
and the treatment of RMC from the $p\mu p$ molecule.

\section{Calculational Method}

We employ heavy-baryon 
chiral perturbation theory  \cite{hbchpt}
and use the effective Lagrangian 
${\cal L}_{\mathrm{ch}}$ 
as given in \cite{bkm95}. 
${\cal L}_{\mathrm{ch}}$ is written in the most general form 
involving pions and heavy nucleons
in external weak- and electromagnetic-fields 
consistent with chiral symmetry.
We expand ${\cal L}_{\mathrm{ch}}$ in increasing chiral order as: 
\begin{equation}
{\cal L}_{\mathrm{ch}} = 
{\cal L}_{\pi}^{(0)} + 
{\cal L}_{\pi N}^{(0)} + {\cal L}_{\pi N}^{(1)} +\cdots \; .
\label{Lag} 
\end{equation}
Here ${\cal L}^{(\bar{\nu})}$ represents
terms of chiral order $\bar{\nu}$ given by
${\bar{\nu}} \equiv d + \half  n - 2$,
where $d$ is the summed power of 
the derivative and the pion mass, and
$n$ denotes the number of nucleon fields involved
in a given term \cite{wei90}. 
We limit ourselves here to 
a next-to-leading chiral order (NLO) calculation
and therefore we only keep terms 
with ${\bar{\nu}} = 0$ and ${\bar{\nu}} = 1$.
To this chiral order we need only consider tree diagrams,
and then ${\cal L}_{\pi N}^{(1)}$ simply represents
$1/M$ ``nucleon recoil'' corrections to 
the leading ``static'' part ${\cal L}_{\pi N}^{(0)}$. 
We give below the explicit expressions 
for the ${\cal L}_{\pi}^{(0)}$, 
${\cal L}_{\pi N}^{(0)}$ and ${\cal L}_{\pi N}^{(1)}$,
in which only  terms of direct relevance for our NLO calculation
are retained.
\begin{eqnarray}
{\cal L}_{\pi}^{(0)} \, &=& \, 
\frac{f_\pi^2}{4} {\mathrm{Tr}} \left [
D_\mu U D^\mu U \right ] \, + \dots \label{lpi0} \\
{\cal L}_{\pi N}^{(0)} \, &=& \,  
{\bar N} \left \{  i v\cdot D + 
g_A S\cdot u \right \}  N \label{lpin0} \\
{\cal L}_{\pi N}^{(1)} \, &=& \, 
{\bar N} \Big \{ \frac{1}{2M} 
(v\cdot D)^2 - \frac{1}{2M} D\cdot D
- \frac{i g_A}{2 M} \{ S\cdot D , v\cdot u \}_{+} 
\nonumber \\
& - & 
\frac{i}{4M} [S^\mu ,S^\nu ]_{-} 
\left ( (1+\kappa_v) f_{\mu\nu}^+ + \half
(\kappa_s - \kappa_v) 
{\mathrm{Tr}}  f_{\mu\nu}^+ \right ) \Big  \} N 
\, + \dots  \; . 
\label{lpin1} \\
\nonumber 
\label{llterms}
\end{eqnarray} 
Here $U = \sqrt{1- \vec{\pi}^2 /f_\pi^2} + i
\vec{\tau}\cdot\vec{\pi} /f_\pi$ denotes 
the chiral field in the sigma gauge, and 
$N$ the heavy nucleon spinor of mass $M$. 
We have also used other standard notations, see \cite{bkm95}: 
\begin{eqnarray}
D_\mu U \, &\equiv&  \, \partial_\mu U - i ({\cal{V}}_\mu  + {\cal{A}}_\mu ) U
+ i U   ({\cal{V}}_\mu  - {\cal{A}}_\mu ) 
\nonumber \\
U &\equiv& u^2 ;  \hspace{8mm} 
u_\mu \equiv i u^\dagger D_\mu U u^\dagger  
\nonumber \\
D_\mu N \, &\equiv& \, \partial_\mu N + \half [u^\dagger,\partial_\mu u ]_{-} N
- \frac{i}{2} u^\dagger ({\cal{V}}_\mu + {\cal{A}}_\mu ) u 
- \frac{i}{2} u ({\cal{V}}_\mu - {\cal{A}}_\mu ) u^\dagger 
\nonumber \\
F_\mu^R &\equiv& {\cal{V}}_\mu + {\cal{A}}_\mu ; \hspace{8mm} 
F_\mu^L \equiv {\cal{V}}_\mu - {\cal{A}}_\mu 
\nonumber \\
F_{\mu\nu}^{L,R} &\equiv& 
\partial_\mu F_\nu^{L,R} - 
\partial_\nu F_\mu^{L,R} -
i [F_\mu^{L,R} , F_\nu^{L,R} ]_-  
\nonumber \\
f_{\mu\nu}^+ &\equiv& u^\dagger F_{\mu\nu}^{R} u +
 u F_{\mu\nu}^{L} u^\dagger
\; . 
\label{definitions}
\end{eqnarray} 
The covariant derivatives above include 
the external vector and axial vector fields,
${\cal V}_\mu = {\cal{V}}_\mu^a \frac{\tau^a}{2}$
and ${\cal A}_\mu = {\cal{A}}_\mu^a \frac{\tau^a}{2}$, 
respectively. 
If we choose the four-velocity $v_\mu$
to be $v^\mu = (1,\vec{0})$,
the spin operator $S^\mu$ of the heavy nucleon 
becomes $S^\mu = (0,\half\vec{\sigma})$. 
The only parameters appearing in the above expressions 
are the pion decay constant, 
$f_\pi$ = 93 ${\mathrm{MeV}}$, 
the axial vector coupling, $g_A = 1.26$, 
and the nucleon isoscalar and isovector 
anomalous magnetic moments, 
$\kappa_s = -0.12$ and $\kappa_v = 3.71$.
Thus,  to the chiral order of our interest,
${\cal L}_{\mathrm{ch}}$ is well determined.
   
\def\Text(#1,#2)[#3]#4{
%
%
\put(#1,#2){\makebox[0pt][#3]{#4}}
}
\vspace{10mm}
\begin{figure}[ht]
\begin{picture}(380,100)
\ArrowLine(50,0)(70,50)
\ArrowLine(70,50)(50,100)
\ArrowLine(150,0)(130,50)
\ArrowLine(130,50)(150,100)
\ZigZag(70,50)(130,50)28
\Photon(60,25)(25,65)44
\Text(40,0)[c]{$\mu^-$}
\Text(40,100)[c]{$\nu$}
\Text(160,0)[c]{$p$}
\Text(160,100)[c]{$n$}
\Text(15,65)[c]{$\gamma$}
\Text(100,55)[c]{$W^-$}
\Text(100,0)[c]{(a)}
\ArrowLine(250,0)(270,50)
\ArrowLine(270,50)(250,100)
\ArrowLine(350,0)(330,50)
\ArrowLine(330,50)(350,100)
\ZigZag(270,50)(295,50)24
\DashLine(295,50)(330,50)4
\Photon(260,25)(225,65)44
\Text(240,0)[c]{$\mu^-$}
\Text(240,100)[c]{$\nu$}
\Text(360,0)[c]{$p$}
\Text(360,100)[c]{$n$}
\Text(215,65)[c]{$\gamma$}
\Text(285,55)[c]{$W^-$}
\Text(315,55)[c]{$\pi^-$}
\Text(300,0)[c]{(b)}
\end{picture}
\caption{}
\vspace{10mm}
\label{fig1}
\end{figure}
\begin{figure}[ht]
\begin{picture}(380,100)
\ArrowLine(50,0)(70,50)
\ArrowLine(70,50)(50,100)
\ArrowLine(150,0)(130,50)
\ArrowLine(130,50)(150,100)
\ZigZag(70,50)(130,50)28
\Photon(140,25)(175,65)44
\Text(40,0)[c]{$\mu^-$}
\Text(40,100)[c]{$\nu$}
\Text(160,0)[c]{$p$}
\Text(160,100)[c]{$n$}
\Text(185,65)[c]{$\gamma$}
\Text(100,55)[c]{$W^-$}
\Text(100,0)[c]{(a)}
\ArrowLine(250,0)(270,50)
\ArrowLine(270,50)(250,100)
\ArrowLine(350,0)(330,50)
\ArrowLine(330,50)(350,100)
\ZigZag(270,50)(295,50)24
\DashLine(295,50)(330,50)4
\Photon(340,25)(375,65)44
\Text(240,0)[c]{$\mu^-$}
\Text(240,100)[c]{$\nu$}
\Text(360,0)[c]{$p$}
\Text(360,100)[c]{$n$}
\Text(385,65)[c]{$\gamma$}
\Text(285,55)[c]{$W^-$}
\Text(315,55)[c]{$\pi^-$}
\Text(300,0)[c]{(b)}
\end{picture}
\caption{}
\vspace{10mm}
\label{fig2}
\end{figure}
\begin{figure}[ht]
\begin{picture}(380,100)
\ArrowLine(50,0)(70,50)
\ArrowLine(70,50)(50,100)
\ArrowLine(150,0)(130,50)
\ArrowLine(130,50)(150,100)
\ZigZag(70,50)(130,50)28
\Photon(140,75)(185,95)44
\Text(40,0)[c]{$\mu^-$}
\Text(40,100)[c]{$\nu$}
\Text(160,0)[c]{$p$}
\Text(160,100)[c]{$n$}
\Text(195,95)[c]{$\gamma$}
\Text(100,55)[c]{$W^-$}
\Text(100,0)[c]{(a)}
\ArrowLine(250,0)(270,50)
\ArrowLine(270,50)(250,100)
\ArrowLine(350,0)(330,50)
\ArrowLine(330,50)(350,100)
\ZigZag(270,50)(295,50)24
\DashLine(295,50)(330,50)4
\Photon(340,75)(385,95)44
\Text(240,0)[c]{$\mu^-$}
\Text(240,100)[c]{$\nu$}
\Text(360,0)[c]{$p$}
\Text(360,100)[c]{$n$}
\Text(395,95)[c]{$\gamma$}
\Text(285,55)[c]{$W^-$}
\Text(315,55)[c]{$\pi^-$}
\Text(300,0)[c]{(b)}
\end{picture}
\caption{}
\vspace{10mm}
\label{fig3}
\end{figure}

\begin{figure}[ht]
\begin{picture}(380,100)
\ArrowLine(150,0)(170,50)
\ArrowLine(170,50)(150,100)
\ArrowLine(250,0)(230,50)
\ArrowLine(230,50)(250,100)
\ZigZag(170,50)(195,50)24
\DashLine(195,50)(230,50)4
\Photon(220,50)(220,100)44
\Text(140,0)[c]{$\mu^-$}
\Text(140,100)[c]{$\nu$}
\Text(260,0)[c]{$p$}
\Text(260,100)[c]{$n$}
\Text(230,100)[c]{$\gamma$}
\Text(185,35)[c]{$W^-$}
\Text(215,40)[c]{$\pi^-$}
\end{picture}
\caption{}
\vspace{10mm}
\label{fig4}
\end{figure}

\begin{figure}[ht]
\begin{picture}(380,100)
\ArrowLine(50,0)(70,50)
\ArrowLine(70,50)(50,100)
\ArrowLine(150,0)(130,50)
\ArrowLine(130,50)(150,100)
\ZigZag(70,50)(130,50)28
\Photon(130,50)(175,85)44
\Text(40,0)[c]{$\mu^-$}
\Text(40,100)[c]{$\nu$}
\Text(160,0)[c]{$p$}
\Text(160,100)[c]{$n$}
\Text(180,80)[c]{$\gamma$}
\Text(100,55)[c]{$W^-$}
\Text(100,0)[c]{(a)}
\ArrowLine(250,0)(270,50)
\ArrowLine(270,50)(250,100)
\ArrowLine(350,0)(330,50)
\ArrowLine(330,50)(350,100)
\ZigZag(270,50)(295,50)24
\DashLine(295,50)(330,50)4
\Photon(330,50)(375,85)44
\Text(240,0)[c]{$\mu^-$}
\Text(240,100)[c]{$\nu$}
\Text(360,0)[c]{$p$}
\Text(360,100)[c]{$n$}
\Text(380,85)[c]{$\gamma$}
\Text(285,55)[c]{$W^-$}
\Text(315,55)[c]{$\pi^-$}
\Text(300,0)[c]{(b)}
\end{picture}
\caption{}
\vspace{10mm}
\label{fig5}
\end{figure}

\begin{figure}[ht]
\begin{picture}(380,100)
\ArrowLine(150,0)(170,50)
\ArrowLine(170,50)(150,100)
\ArrowLine(250,0)(230,50)
\ArrowLine(230,50)(250,100)
\ZigZag(170,50)(195,50)24
\DashLine(195,50)(230,50)4
\Photon(195,50)(195,100)44
\Text(140,0)[c]{$\mu^-$}
\Text(140,100)[c]{$\nu$}
\Text(260,0)[c]{$p$}
\Text(260,100)[c]{$n$}
\Text(205,100)[c]{$\gamma$}
\Text(185,35)[c]{$W^-$}
\Text(215,40)[c]{$\pi^-$}
\end{picture}
\caption{}
\vspace{10mm}
\label{fig6}
\end{figure}
We consider all possible Feynman diagrams 
up to chiral order $\nu$ = 1 
which contribute to the process 
$\mu^- + p \to n + \nu + \gamma$.
These are displayed in Figs.\ref{fig1}-\ref{fig6}. 
The zigzag lines in these diagrams represent
the $W^-$ boson that couples to the leptonic 
and hadronic currents in the standard manner.
In the actual calculation, 
taking the limit $m_W \to \infty$,
we make the substitution: 
$W_\mu^- \to ({\cal{V}}_\mu^- - {\cal{A}}_\mu^-) 
(\tau^1 - i \tau^2)/2$,
and treat $\cal{V}$ and $\cal{A}$ 
as static external vector and axial sources, respectively. 
Then the diagrams in Figs.\ref{fig1}-\ref{fig6}
reduce to those that would result from
the simple current-current interaction of the $V-A$ form.
The reason for explicitly retaining the $W^-$ boson lines
is to clearly separate the different photon vertices
(see e.g. Fig.\ref{fig6}). 
The leptonic vertices in these Feynman diagrams are 
of course well known.
The hadronic vertices are obtained 
by expanding the ChPT Lagrangian
[Eqs. (\ref{Lag}), (\ref{lpi0}), (\ref{lpin0})
and (\ref{lpin1})]
in terms of the elementary fields $N$, $\pi$, 
$\cal{V}$ and $\cal{A}$ and their derivatives. 
The leading order terms arise from 
${\cal L}_{\pi N}^{(0)}$, whereas the NLO 
contributions are $1/M$ ``recoil'' corrections due to 
${\cal L}_{\pi N}^{(1)}$. 
The evaluation of the transition amplitudes 
corresponding to these Feynman diagrams 
is straightforward. 
We denote by $M_i$ ($i=1\dots 6$) 
the invariant transition amplitudes
corresponding to Fig.(\ref{fig1})-({\ref{fig6}), respectively. 
They are given by:
\vspace{-0.2cm}
\begin{eqnarray}
M_1 \, &=& \, \epsilon^\beta (\lambda) 
\left [ \bar{u}_\nu (s) \gamma_\tau (1-\gamma_5 ) 
\frac{\FMslash{\mu} -\FMslash{k} + m_\mu } 
{ 2 (k\cdot\mu ) } \gamma_\beta u_\mu ( s^\prime ) \right ] \;
\left [ H_n^\dagger (\sigma)  h_1^\tau H_p (\sigma^\prime ) \right ]
\label{m1} \\
M_i \, &=& \, \left [ \bar{u}_\nu (s) \gamma_\tau 
(1-\gamma_5 ) u_\mu ( s^\prime ) \right ] \;
\left [ H_n^\dagger (\sigma)  h_i^\tau (\lambda) H_p 
(\sigma^\prime ) \right ]\,,  \; i=2,3,4,5,6 \; , 
\label{mi} \\
\nonumber
\end{eqnarray}
\vspace{-0.3cm}
which include the following hadronic operators:
\begin{eqnarray}
h_1^\tau \, &=& \,
\left [ (v^\tau - 2g_A S^\tau) + 2 g_A \frac{(q_L)^\tau} 
{(q_L)^2 - m_\pi^2} (S \cdot q_L ) \right ] \nonumber \\
&+& \Biggl \{ \frac{1}{2M} \left [ (p+n)^\tau - v^\tau v\cdot(p+n) 
\right ] - \frac{1}{M} (1+\kappa_v) i 
\epsilon^{\mu\tau\nu\alpha} (q_L)_\mu v_\nu S_\alpha \nonumber \\
&+& \frac{g_A}{M} v^\tau S\cdot(p+n)
- \frac{g_A}{M} S\cdot(p+n)
\frac{(q_L)^\tau}{q_L^2 - m_\pi^2}  (v\cdot q_L) \Biggr \}
\label{h1}
\end{eqnarray}
\vspace{-0.7cm}
\begin{eqnarray}
h_2^\tau (\lambda) \, &=& \,
\frac{1}{2M}
\frac{1}{E_p - \omega_k} 
\left [ (v^\tau - 2g_A S^\tau) + 2 g_A \frac{(q_N)^\tau} 
{(q_N)^2 - m_\pi^2} (S \cdot q_N ) \right ]
\nonumber \\ &\cdot& 
\left [ \epsilon(\lambda)  \cdot (2p-k) + (2+\kappa_s + 
\kappa_v)(-i) \epsilon_{\alpha\beta\gamma\rho} 
\epsilon^\alpha (\lambda) k^\beta v^\gamma S^\rho \right ]
\label{h2}
\end{eqnarray}
\vspace{-0.7cm}
\begin{eqnarray}
h_3^\tau (\lambda) \, &=& \,
\frac{1}{2M}
\frac{1}{E_n + \omega_k} 
\left [ (v^\tau - 2g_A S^\tau) + 2 g_A \frac{(q_N)^\tau} 
{(q_N)^2 - m_\pi^2} (S \cdot q_N ) \right ]
\nonumber \\ &\cdot& 
\left [ (\kappa_s - \kappa_v)(-i) \epsilon_{\alpha\beta\gamma\rho} 
\epsilon^\alpha (\lambda) k^\beta v^\gamma S^\rho \right ]
\label{h3}
\end{eqnarray}
\vspace{-0.7cm}
\begin{eqnarray}
h_4^\tau (\lambda) \, = \,
(&-&) \frac{(q_N)^\tau (2q_L +k)\cdot\epsilon(\lambda)} 
{(q_N^2-m_\pi^2)(q_L^2-m_\pi^2)} 
\nonumber \\
&\cdot&
\left [2g_A (S\cdot q_L ) - \frac{g_A}{M} S\cdot(p+n) 
(q_L \cdot v) \right ]
\label{h4}
\end{eqnarray}
\vspace{-0.7cm}
\begin{eqnarray}
h_5^\tau (\lambda) \, = \,
&2& g_A \frac{(q_N)^\tau}{(q_N)^2 - m_\pi^2} 
(S \cdot \epsilon(\lambda) )
+ \frac{g_A}{M} \frac{(q_N)^\tau}{(q_N)^2 - m_\pi^2} 
(v\cdot q_N) (S\cdot\epsilon(\lambda)) 
\nonumber \\
&-& \frac{g_A}{M} (S\cdot\epsilon(\lambda)) v^\tau 
 - \frac{\epsilon^\tau (\lambda)}{2M} + 
\frac{1}{2M} (1+\kappa_v) i \epsilon^{\tau\alpha\beta\rho}
\epsilon_\alpha (\lambda) v_\beta S_\rho 
\label{h5}
\end{eqnarray}
\vspace{-0.7cm}
\begin{equation}
h_6^\tau (\lambda) \, = \,
\frac{\epsilon^\tau (\lambda)}{q_L^2 - m_\pi^2} 
\left [ 2 g_A (S\cdot q_L ) - \frac{g_A}{M} 
S\cdot (p+n)(v\cdot q_L) \right ]
\, .
\label{h6}
\end{equation}
In these expressions,
$\mu$, $\nu$, $p=(E_p,\vec{p})$, 
$n=(E_n,\vec{n})$ and $k=(\omega_k,\vec{k})$ 
are the four-momenta of the muon, neutrino, 
proton, neutron and photon, respectively. 
The $z$-components of the spins 
of the muon, neutrino, 
proton and neutron are denoted by 
$s$, $s^\prime$, $\sigma^\prime$ and $\sigma$, 
respectively, while $\epsilon(\lambda)$ 
stands for the photon polarization vector. 
We have also defined 
$q_L = n-p$ and $q_N = n-p+k$.

The pion-pole diagrams, 
Figs.\ref{fig1}(b), \ref{fig2}(b), \ref{fig3}(b), \ref{fig4}, 
\ref{fig5}(b) and \ref{fig6}, 
originate from 
${\cal L}_\pi^{(0)}$, Eq.(\ref{lpi0}). 
The coupling of the axial vector 
to the $\pi$ generates these Feynman diagrams. 
In ChPT the pion-pole contributions,
which arise automatically 
from a well-defined chiral Lagrangian,
are completely determined by the chiral Lagrangian.
The fact that they need not be put in by hand
constitutes a major advantage of the ChPT approach 
over the phenomenological approaches which have
been used in the earlier calculations \cite{bf87,mw59,opa64}. 
For example, the term originating from Fig.5(b) 
does not appear in  Ref.\cite{opa64}. In addition, 
due to the pure pseudoscalar pion 
nucleon coupling, the pion-pole terms are proportional 
to $1/M$ in Ref.\cite{opa64}. 
In this context it is also worthwhile to mention 
that the pseudoscalar coupling $g_P$ itself
does not appear explicitly in ChPT calculations of 
the transition amplitudes since $g_P$ 
is effectively accounted for via the pion-pole diagrams.
As mentioned in the introduction,
${\cal L}_{\mathrm{ch}}$ determines $g_P$ 
\cite{bkm94,feaetal97}.
However, 
since the same ${\cal L}_{\mathrm{ch}}$ directly 
determines the transition amplitude of RMC,
$g_P$ does not feature in our expressions 
for $M_i$'s.

It is safe to assume both the muon and the proton 
to be at rest by neglecting the binding and kinetic energies 
of the $\mu p$ atom.
Thus, $\mu =(m_\mu,\vec{0})$ and $p=(M,\vec{0})$. 
For the neutron four-momentum $n$,
we retain its three-momentum $\vec{n}$ but
neglect the recoil energy, or
$E_n = M + \vec{n}^2 /2 M \approx M$. 
The maximal value of $|\vec{n}|$ equals $m_\mu$ 
giving a recoil energy $\vec{n}^2 /2 M \approx $ 6 MeV,
which is small even compared with $m_\mu$. 
With $n \approx (M, \vec{n})$, we have 
$q_L = (0,\vec{n})$ and $q_N =(\omega_k , \vec{n}+\vec{k})$.
Consequently, all terms proportional 
to $v\cdot q_L$ vanish. 
We choose to work in the Coulomb gauge 
with the result  $v \cdot \epsilon(\lambda) =0$. 
With this gauge choice and 
the above kinematical approximations,
the hadronic radiation diagrams, 
Figs.(\ref{fig2}) and (\ref{fig3}), 
become ${\cal{O}}(1/M^2)$
[see Eqs. (\ref{h2}) and (\ref{h3})],
and therefore do not contribute 
to the chiral order under consideration. 
Moreover, in the sum $h_2 + h_3$, 
the terms proportional to $\kappa_v$
vanish in our approach (to the order under consideration), 
whereas in the treatment of, e.g., \cite{mw59},
these terms are numerically large. 

\section{Numerical Results}

As stated, we consider here only the RMC 
from the $\mu p$ atomic state with the 
hyperfine states unseparated.
Within our kinematical approximations
the spin-averaged total capture rate is given by
\begin{eqnarray}
\Gamma_{\mathrm{RMC}} \,  = \,  &&
\left ( 
\frac{eG}{\sqrt{2}} \right )^2
\vert \Phi (0) \vert^2
\,
\quart
\,
(2\pi)^4
\,
\int \frac{d^3 n}{(2\pi)^3}
\,
\int \frac{d^3 \nu}{(2\pi)^3}
\,
\int \frac{d^3 k}{(2\pi)^3}
\,
\frac{1}{2 \omega_k}
\nonumber \\ 
&& \times
\delta^{(4)} (n+\nu +k - p - \mu)
\sum_{\sigma{\sigma^\prime}s{s^\prime}\lambda} 
\vert M \vert^2 \; ,
\label{rmc1}
\end{eqnarray}
where the sum is over all spin 
and polarization orientations, 
$M = \sum_{i=1}^6 M_i$,
with $M_i$ given by Eqs. (\ref{m1}) and (\ref{mi});
$\Phi(0)$ is the value of the $\mu p$ atomic wavefunction 
at the origin.
In the kinematical  approximation stated earlier,
Eq.(\ref{rmc1}) simplifies as 
\begin{eqnarray}
\Gamma_{\mathrm{RMC}} \, = \,
\left ( 8\pi^2{\cal{C}} \right )
&&
\int\limits_{0}^{(\omega_k)_{\mathrm{max}} 
\approx m_\mu}
d \, \omega_k \, \omega_k \, (m_\mu - \omega_k )^2
\nonumber \\ 
&&
\times \int 
d \cos \theta 
\sum_{i,j=1,4,5,6}
\; \; 
\sum_{\sigma{\sigma^\prime}s{s^\prime}\lambda} 
\left ( M_i M_j^* \right )_{\vec{n}= -(\vec{\nu} 
+\vec{k}) \atop 
\vert \vec{\nu} \vert = m_\mu - \omega_k } \; ,
\label{rmc2}
\end{eqnarray}
where we have introduced the abbreviation
${\cal{C}} = (eG/\sqrt{2})^2\,(1/2^3\pi^5)$.
The evaluation of the spin sum is tedious but straightforward;
the resulting lengthy expressions will be given elsewhere. 

Table \ref{table1} summarizes the numerical values for the total capture rate   
$\Gamma_{\mathrm{RMC}}$.
We show in the table the breakdown of $\Gamma_{\mathrm{RMC}}$
into the leading-order contribution ${\cal{O}} ((1/M)^0$),
coming from ${\cal{L}}_{\pi N}^{(0)}$,
and the next-to-leading-order contribution ${\cal{O}} (1/M)$, 
arising from ${\cal{L}}_{\pi N}^{(1)}$. 
We also show the value of $\Gamma_{\mathrm{RMC}}$ 
which would result if all the diagrams containing the pion-pole,  
Fig.\ref{fig1}(b), \ref{fig4}, \ref{fig5}(b) and \ref{fig6}, are omitted.
Our result for the total capture rate 
$\Gamma_{\mathrm{RMC}} = 0.075 s^{-1}$
is close to the value given in \cite{opa64}, 
$\Gamma_{\mathrm{RMC}}\,=\,0.069 s^{-1}$, 
and practically identical to 
$\Gamma_{\mathrm{RMC}}\,=\,0.076 s^{-1}$
reported in \cite{fea80}.
Our  ${\cal{O}} (1/M)$ recoil corrections 
account for about $20\%$ 
of the leading order 
${\cal{O}} ((1/M)^0)$ contribution,
which indicates a reasonable convergence 
of the chiral expansion.      
It should be noted that the size of the $1/M$ corrections 
is noticeably larger in the approach of \cite{fea80}. 
As one can see from Table \ref{table1}, about $30\%$ 
of the total value of $\Gamma_{\mathrm{RMC}}$ 
comes from the pion-pole exchange diagrams.

\begin{table}[ht]
\caption{Total RMC capture rate in $s^{-1}$ }
\begin{tabular}{|c|c|c|}
\hline
 & $\Gamma_{\mathrm{RMC}}$ & $\Gamma_{\mathrm{RMC}} 
\vert_{\mathrm{without }\pi}$
\\ \hline
${\cal{O}} ((1/M)^0)$ & $0.061 \; $ & $0.043 \; $
\\ \hline
${\cal{O}} (1/M)$ & $0.014 \; $ & $0.010 \; $
\\ \hline
total & $0.075 \; $ & $0.053 \; $
\\ \hline
\end{tabular}
\label{table1}
\end{table}

In Fig.\ref{fig7} we plot the spectrum of the emitted photons
$d \Gamma_{\mathrm{RMC}} (\omega_k) / d\omega_k$.
In addition to the result of the full calculation,
the figure includes the spectrum corresponding to 
the leading-order calculation, i.e.,
the ${\cal{O}} ((1/M)^0)$ contribution only.
For the sake of comparison, we also show the result of
\cite{bf87,fea80} corresponding to the use of
the Goldberger-Treiman value $g_P = 6.6 g_A$.

\begin{figure}[b]
\psfig{file=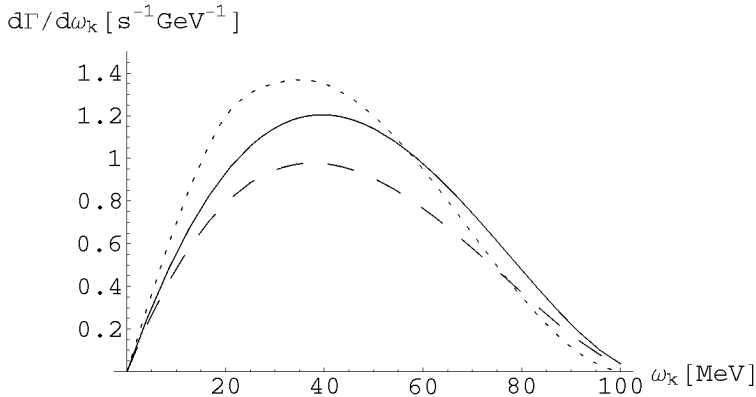,width=4in}
\caption{Spectrum of the emitted photons.
The full line represents the full calculation including 
${\cal{O}}((1/M)^0)$ and ${\cal{O}}((1/M)^1)$;
the dashed line represents the result that contains
only  the ${\cal{O}}((1/M)^0)$ contributions;
the dotted line shows the result of \cite{bf87,fea80} 
with $g_P = 6.6 g_A$.}
\label{fig7}
\end{figure}

\section{Discussion and Conclusions}

A direct comparison of our calculation
with the experimental data \cite{jonetal96}
is premature because we have not considered capture
from the singlet and triplet hyperfine states separately,
or capture from the $p\mu p$ molecular state.
This also means that at this stage we cannot 
directly address the ``$g_P$ problem'' that arose 
from the TRIUMF data \cite{jonetal96}.
However, it is worthwhile to make 
the following remark.
As one can see from Fig.\ref{fig7}
our ChPT calculation gives 
for the spin-averaged $\mu p$-atomic RMC 
a photon spectrum
that is slightly harder 
(by about 10\% for $E_\gamma > $ 60 MeV) 
than what was obtained in 
\cite{bf87,fea80} with the use of
the G-T value, $g_P = 6.6 g_A$.
Meanwhile, as mentioned earlier, 
ChPT gives a value of $g_P$ 
consistent with $g_P = 6.6 g_A$ \cite{bkm94,feaetal97}.
Thus, there is a possibility that,
even with the same value of $g_P$,
a ChPT calculation gives a somewhat 
harder $\gamma$ spectrum 
than the conventional method.
It remains to be seen to what extent
such a difference in the $\gamma$ spectrum
influences the $g_P$ value deduced
from the experimental spectrum in the higher energy region.
Of course, a more quantitative statement can be made 
only after a more detailed ChPT calculation becomes available 
in which the hyperfine states are separated and 
the $p\mu p$-molecular absorption is evaluated.
We also remark that using relativistic kinematics, 
instead of the kinematic approximation employed in Eq.(16), 
softens the photon spectrum to a certain extent.

We repeat that the present calculation
includes only up to the next-to-leading chiral order (NLO) 
contributions.
The next-to-next-to-leading order (NNLO) calculations
are obviously desirable. 
For this one must include the $\bar{\nu} =2$ chiral Lagrangian,
${\cal{L}}_{\pi}^{(2)}$ and ${\cal{L}}_{\pi N}^{(2)}$,
and also loop corrections arising from ${\cal{L}}_{\pi}^{(0)}$ 
and ${\cal{L}}_{\pi N}^{(0)}$. 
The finite contributions from the loop diagrams 
would give momentum-dependent vertices,
which would correspond to the form factors
in the language of the phenomenological 
approach \cite{bf87,mw59,fea80}.
These contributions are probably small but 
it would be reassuring 
to check that explicitly.
One problem in extending the present calculation 
to the next order is that, 
although the forms of ${\cal{L}}_{\pi}^{(2)}$ 
and ${\cal{L}}_{\pi N}^{(2)}$ 
have been determined \cite{gl84,eckmoj}, 
some coefficients of the counter terms 
in ${\cal{L}}_{\pi N}^{(2)}$
still remain undetermined.
On the other hand, chiral expansion for muon capture is
characterized by the expansion parameter $m_\mu /M$,
and is expected to converge reasonably rapidly.
Indeed, in the case of OMC, where the $\nu=2$ calculation
is much less involved,
explicit evaluations \cite{bkm94,feaetal97} show 
that the NNLO contributions amount only to a few percents.
It is likely that, in the case of RMC as well, 
NNLO corrections modify our results only by a few percents. 
In this connection we also note that
the formalism of Bernard et al.\cite{bkm95} used here
does not contain the explicit $\Delta$ degree of freedom
in contrast to the approaches of \cite{hbchpt}.
Although it is desirable to examine 
the importance of the $\Delta$,
we relegate that to future studies.
[After the completion of the present work we learned 
of the first attempt at an NNLO calculation 
by Ando and Min \cite{am97}.]

\begin{ack}
This work is supported in part by the National Science Foundation, 
Grants \# PHY-9319641 and \# PHYS- 9602000. 
\end{ack}

\end{document}